\documentclass[useAMS,usenatbibi,usegraphicx]{mn2e}
\usepackage{psfig}

\newcounter{species} 
\def\ion#1#2{\setcounter{species}{#2}#1$\;${\scriptsize\Roman{species}}\relax}
\def\lion#1#2#3{\setcounter{species}{#2}#1$\;${\scriptsize\Roman{species}}$\;\lambda$#3\relax}
\def\llion#1#2#3{\setcounter{species}{#2}#1$\;${\scriptsize\Roman{species}}$\;\lambda\lambda$#3\relax}

\newcommand{\be}{\begin{equation}}
\newcommand{\ee}{\end{equation}}
\newcommand{\ba}{\begin{eqnarray}}
\newcommand{\ea}{\end{eqnarray}}

\newcommand{\Ms}{M_{\odot}}

\newcommand{\bml}{\begin{mathletters}}
\newcommand{\eml}{\end{mathletters}}

\def\ltsima{$\; \buildrel < \over \sim \;$}
\def\simlt{\lower.5ex\hbox{\ltsima}}
\def\gtsima{$\; \buildrel > \over \sim \;$}
\def\simgt{\lower.5ex\hbox{\gtsima}}
\def\gsim{ \lower .75ex \hbox{$\sim$} \llap{\raise .27ex \hbox{$>$}} }
\def\lsim{ \lower .75ex\hbox{$\sim$} \llap{\raise .27ex \hbox{$<$}} }
\def\msun{\,{\rm M_\odot}}

\title[GW and electo-magnetic emission from tidally disrupted WDs]
{Observing white dwarfs orbiting massive black holes in the gravitational wave 
and electro-magnetic window}

\author[A. Sesana et al.]{A. Sesana$^{1}$, A. Vecchio$^{2}$, M.~Eracleous$^{1,3}$ and S.~Sigurdsson$^{1,3}$\\
$^{1}$ Center for Gravitational Wave Physics, The Pennsylvania State University, University Park, PA 16802, USA\\
$^{2}$ School of Physics and Astronomy, University of Birmingham, 
Edgbaston, Birmingham, B15 2TT, UK\\
$^{3}$ Department of Astronomy \& Astrophysics, The Pennsylvania State University, University Park, PA 16802, USA}

\begin{document}

\date{Received ---}

\maketitle

\begin{abstract}

We consider a potentially new class of gravitational wave sources
consisting of a white dwarf coalescing into a massive black hole in
the mass range $\sim 10^4-10^5\,\msun$. These sources are of
particular interest because the gravitational wave signal produced during
the inspiral phase can be detected by the 
{\it Laser Interferometer Space Antenna} ({\it LISA}) 
and is promptly followed, in an extended portion of the black hole and white 
dwarf mass parameter space, by an electro-magnetic signal generated by 
the tidal disruption of the star, detectable with X-ray, optical and UV 
telescopes. This class of sources could therefore yield a
considerable number of scientific payoffs, that include 
precise cosmography at low redshift, 
demographics of black holes in the mass range $\sim 10^4 - 10^5\Ms$,
insights into dynamical interactions and populations of white dwarfs 
in the cores of dwarf galaxies, as well as a new probe into the
structure and equation of state of white dwarfs.
By modelling the gravitational and electromagnetic radiation produced by these
events, we find them detectable in both observational 
windows at a distance $\approx 200$ Mpc, and possibly
beyond for selected regions of the parameter space. We also estimate the
detection rate for a number of model assumptions about black hole and 
white dwarf mass functions and dynamical interactions:  the rate is 
(not surprisingly) highly uncertain, ranging from 
$\sim 0.01\,\mathrm{yr}^{-1}$ to $\sim 100\,\mathrm{yr}^{-1}$.
This is due to the current limited theoretical understanding
and minimal observational constraints for these objects and processes. 
However, capture rate scaling arguments favor the high end of the above 
range, making likely the detection of several events during the {\it LISA} 
lifetime. 
\end{abstract}

\section{Introduction}

The simultaneous detection of sources in both the electro-magnetic
band - which provides a measurement of the source redshift, $z$ -- and
the gravitational wave (GW) window -- which yields a direct
determination of the luminosity distance $D_\mathrm{L}$ to the source
-- could revolutionize cosmography by determining the distance scale
of the Universe in a precise, calibration-free way. This was pointed
out initially by Schutz (1986) in the context of ground-based
observations of GWs from coalescing compact binaries with the network
of ground-based laser interferometers now in operation (Whitcomb 2008). 
The observational capability of space-based instruments
such as the {\it Laser Interferometer Space Antenna} ({\it LISA};
Bender et al. 1998), which could observe many sources at high
signal-to-noise ratio (SNR) and large redshift, has attracted much
attention recently. Several scenarios have been considered, primarily
related to the identification of the host galaxy or galaxy cluster
of massive black hole (MBH) binary systems detected in GWs (Cutler
1998; Hughes 2002; Menou 2003; Vecchio 2004; Holz \& Hughes 2005; Lang \& Hughes
2006, 2008; Kocsis et al. 2006, 2007a,b; Arun et al. 2007; 
Cornish \& Porter 2008; Trias \& Sintes 2008), 
and the possible electro-magnetic signatures produced by the
pre-glow/afterglow of the MBH mergers (Milosavljevi{\'c} \& Phinney 2005; 
Dotti et al. 2006). The main obstacles to such groundbreaking observations 
are either the possible paucity of sources likely to produce significant 
gravitational and electro-magnetic radiation detectable to cosmological 
distances and/or the rather poor angular resolution of GW instruments 
(e.g. Cutler 1998; Hughes 2002; Vecchio 2004; Lang \& Hughes
2006; Arun et al 2007; Cornish \& Porter 2008; Trias \& Sintes 2008), 
which could inhibit the electro-magnetic identification 
of the host.

In this paper, we discuss a new class of GW sources that have received
little attention so far (Menou, Haiman \& Kocsis 2008): the inspiral
of a white dwarf (WD) around a MBH in the mass range $\sim
10^4 - 10^5\Ms$ followed by the tidal disruption of the star before 
it plunges into
the MBH. As we will show, these sources may be observable at low
redshift (a few hundreds Mpc) with {\it LISA} \emph{and} their electro-magnetic
emission may be detectable with X-ray observatories and optical ground
based telescopes. From the GW point of view, a MBH-WD binary is a different
flavour of the so-called Extreme-Mass Ratio Inspirals or EMRIs. Traditionally, 
the fiducial EMRI is taken to be a stellar-mass $\sim 10\,\msun$ black
hole orbiting a $10^6\,\msun$ MBH (Barack \& Cutler 2004). The key
difference between a "traditional EMRI" and a MBH-WD system considered in this
paper is that for a range of MBH and WD masses, the inspiral 
does not proceed all 
the way until the compact object falls into the MBH horizon, 
but it terminates with the
tidal disruption of the WD producing an electro-magnetic 
signature (for a MBH mass
$\simgt 3\times 10^5\,\msun$, a WD survives throughout the 
whole inspiral and the 
system behaves just like a traditional EMRI). 
The observation of both gravitational and electro-magnetic
signals from the same source provides a direct and calibration-free
measurement of the $D_L(z)$ relationship and opens new avenues for
cosmography and, more directly (due to the low redshift of most of the
expected sources) a completely independent determination of the Hubble
parameter $H_0$ that does not depend on any distance calibration. 
This class of sources can also provide new insights
into a number of unanswered questions in relativistic astrophysics: (i)
the demographics of MBHs in the mass range $\sim 10^4 - 10^5\Ms$ --
only a handful of MBH candidates with masses below $10^6\msun$ (Greene
\& Ho 2004; Barth, Greene \& Ho 2005) is known to date,
their mass estimate is rather uncertain, since it is based on 
the emission-line spectra of the active nuclei, and none of them have 
masses below $10^5\msun$ -- (ii) the populations of WDs and the dynamical
processes that take place in the cores of dwarf galaxies, that are unknown and 
unconstrained by observations, and 
(iii) the structure and equation of state of WDs -- the
exact point at which tidal disruption occurs indeed depends on the WD
equation of state (e.g. Magorrian \& Tremaine 1999), and the
electro-magnetic signature carries information about the WD
composition.

The paper is organized as follows: in Section~\ref{s:GW} we identify
the mass range of WDs and central MBHs that lead to the tidal disruption
of the star before it plunges onto the black hole and we 
determine the volume of the Universe
that {\it LISA} will be able to survey; in Section~\ref{s:rate} we
derive the GW detection rate of these sources and discuss its uncertainties; in
Section~\ref{e:EM} we model the electro-magnetic counterpart to MBH-WD
EMRIs; finally, in Section~\ref{e:summary} we summarize the main
results and our conclusions.

\section{The gravitational wave signal}
\label{s:GW}

\subsection{Mass parameter space}
Let us consider a MBH of mass $M$ and an inspiralling WD of mass $m$
and radius $r$.  The tidal disruption radius of the WD is:
\begin{equation}
R_{\rm td}=\left(\eta^2\frac{M}{m}\right)^{1/3}r\,,
\label{rtd}
\end{equation}
where $\eta$ is a factor of order unity that depends on the star
equation of state (for sake of simplicity we set $\eta=1$ in our
analysis). $R_{\rm td}$ can be expressed as a function of only $m$ and
$M$ once the WD $m(r)$ relation is fixed. Here we use a polytrope
approximation with $\Gamma=5/3$ (Nauenberg 1972):
\begin{equation}
r=8\times10^8{\rm cm}\left(\frac{M_{\rm CH}}{m}\right)^{1/3}{\mathcal F}^{3/4}(m),
\label{rwd}
\end{equation}
where ${\mathcal F}(m)=1- (m/M_{\rm CH})^{4/3}$ and $M_{\rm CH}\simeq
1.44\msun$ is the Chandrasekhar mass. There are two characteristic
length scales (and associated frequencies) that one needs to consider
with respect to the tidal radius. The first is the orbital separation
$R_\mathrm{LSO}$ corresponding to the \emph{last stable orbit} where
the binary encounters a dynamical instability and there is a
transition from the adiabatic inspiral to the freely falling plunge;
for an inspiral onto a Schwarzschild MBH $R_\mathrm{LSO}=
M(6+2e_f)/(1+e_f)$, where $e_f$ is the eccentricity of
the orbit at the plunge, and we have adopted geometrical
units, in which $c = G = 1$. The second is the separation $R_\mathrm{H}$ 
at which the
WD falls into the MBH horizon, or, in other words, the merged binary
forms a common horizon and the system settles down through the ringing
of a Kerr MBH. For a Schwarzschild MBH $R_\mathrm{H}=2 M$.

We can therefore distinguish the following three  cases:
\begin{enumerate}
\item $R_{\rm td}>R_\mathrm{LSO}$: 
the star is disrupted
before the last stable orbit. The in-spiral is interrupted generating
a distinct electro-magnetic signature observable both in X-ray and in
Optical/UV that we will describe in detail in Section \ref{e:EM}.
\item $R_\mathrm{H} < R_{\rm td} < R_{\rm LSO}$: the star is disrupted
during the merger phase but before it falls into the BH horizon:
the inspiral phase is essentially unchanged with respect to the "standard
EMRI" evolution. Whether an object disrupted while plunging would
leave any identifiable electro-magnetic signature is an open question
that we will not try to address in this work. In the following we will
derive event rates with and without including this possibility.
\item $R_\mathrm{H} > R_{\rm td}$: the dynamical
evolution is indistinguishable from the "traditional" EMRIs considered 
in the literature so far, and the only trace that the compact object is a 
WD is entirely contained in the measurement of the 
compact object mass through the observation of the GW signal
evolution. No electro-magnetic signal is expected.
\end{enumerate}

Our analysis is focused on events that may produce an electro-magnetic
counterpart, through the tidal disruption of a WD; we will therefore
restrict our attention to the scenarios (i) and (ii). In the following
we will start by considering a WD orbiting a Schwarzschild MBH; we
discuss the impact of spins on the results in Section 2.3.
Setting $R_{\rm td}$ equal to $R_\mathrm{LSO}$ and $R_\mathrm{H}$, respectively, we
find the two critical black masses that mark the transition between
the three cases: 
\be 
M \simeq
3.3\times10^4\msun\left(\frac{1+e_f}{1+e_f/3}\right)^{3/2}
\left(\frac{m}{\msun}\right)^{-1}{\mathcal F}^{3/4}(m),
\label{Mm1}
\ee
\be
M \simeq  1.7\times10^5\msun \,\left(\frac{m}{\msun}\right)^{-1}\,
{\mathcal F}^{3/4}(m).
\label{Mm2}
\ee 
Clearly, the more massive the MBH, the lighter the WD in order
to have tidal disruption 
before the star plunges into the MBH horizon. Surveys of
this class of EMRIs will therefore probe the MBH mass function in the
range $\sim 10^4 - 10^5\,\msun$. Notice that neutron stars are
disrupted by black holes with masses of the order of tens of solar
masses: gravitational radiation is emitted in the frequency window 
accessible through ground-based laser interferometers (e.g. Vallisneri 2000)
and we will therefore not consider these sources in this paper.

\subsection{Observable volume}

\begin{figure}
\centerline{\psfig{file=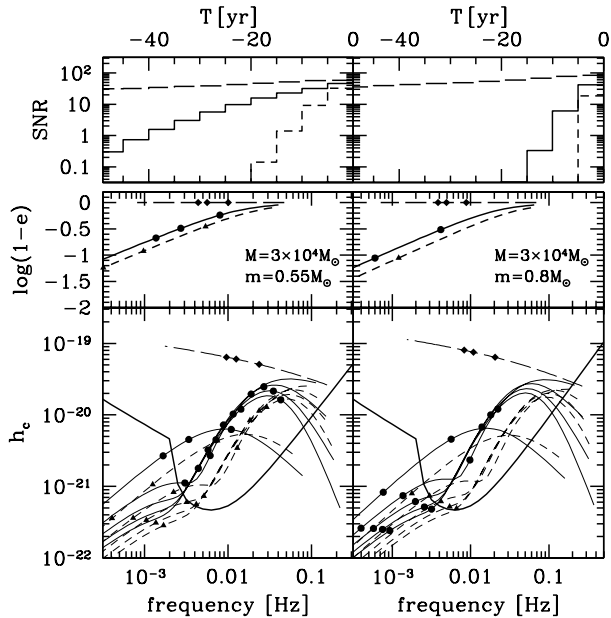,width=84.0mm}}
\caption{SNR build-up during the last years of inspiral for two
different WD and MBH mass combinations at a luminosity distance of 
100 Mpc. In each panel, long--dashed
lines are for a WD with $e_f=0$ (i.e. in circular orbit), solid lines
are for $e_f=0.1$ and short--dashed lines are for $e_f=0.2$. {\it
Lower panels}: characteristic amplitude $h_c$ of the signal versus 
observed frequency 
(for eccentric binaries the first five harmonics are plotted). The
points mark, from right to left, the frequency emitted at 1, 5 and 10
yr before the WD reaches the maximum between $R_{\rm td}$ and $R_{\rm LSO}$. 
The thick solid line is the {\it LISA} sensitivity with the galactic 
WD-WD confusion noise
(according to the prescription given by BC04) taken into account.  {\it Middle
panels}: eccentricity of the binary versus the orbital frequency. {\it
Upper panels}: SNR integrated in different 5-yr observation bins
before the final coalescence.}
\label{fig2}
\end{figure}

\begin{figure}
\centerline{\psfig{file=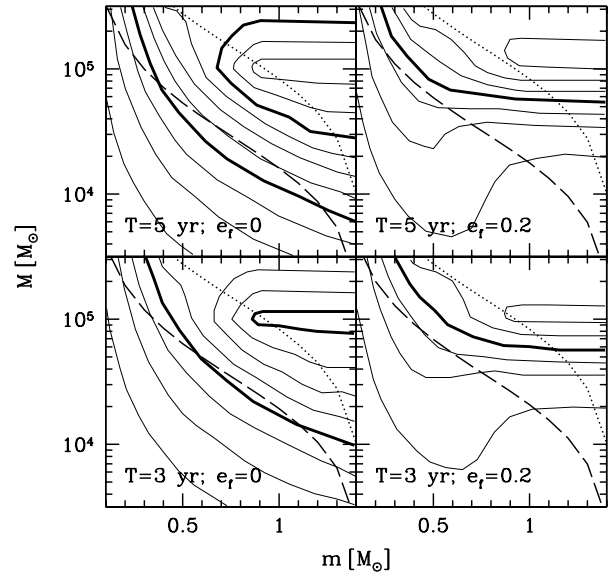,width=84.0mm}}
\caption{Contour plot of $D_{\rm MAX}$ assuming a detection threshold
corresponding to SNR$=30$ and for two different choice of observation 
time -- $T=3$ and 5~yr -- and final eccentricity,  $e_f=0$ and 0.2.  
In each panel, the thin contours are
placed every 50 Mpc with distance increasing from the bottom-left to
the top-right; the thick contours label 200 Mpc and 400 Mpc.
MBH-WD binaries below the dashed line result in the WD disruption before the
last stable orbit, and an electro-magnetic counterpart is expected
(see Section \ref{e:EM}). 
For MBH-WD systems between the dashed and the dotted
lines, the WD is disrupted during the final plunge, {\it before}
crossing the MBH event horizon. }
\label{fig5}
\end{figure}

Having identified the region of interest in the mass plane, we can now 
compute the volume of the Universe that will be accessible to {\it LISA}. We
begin by computing the (angle-averaged) SNR at which {\it LISA} will be 
able to observe
a system at a fiducial luminosity distance that we set to $D_L =
100\,$ Mpc. The SNR scales as $1/D_L$ and, once we specify the minimum
SNR needed for detection, it is straightforward to quantify the depth
of {\it LISA} surveys as a function of the source parameters. We
model the EMRI waveforms according to Barack \& Cutler (2004;
hereafter BC04), therefore adopting what are known as 
``analytic kludge'' waveforms:
the orbits are instantaneously approximated as Newtonian ellipses 
and gravitational radiation 
is given by the corresponding Peters and Mathews formula 
(Peters \& Mathews 1963; Peters 1964), 
but perihelion direction, orbital plane, semi-major axis and 
eccentricity evolve according to 
post-Newtonian equations. The signal is generated by integrating backward in
time Equations (28--30) of BC04 starting from the orbital frequency
emitted at the minimum allowed orbital separation. The latter is set to be
the larger between the last stable orbit and the tidal disruption radius 
of the WD. The
angle-averaged SNR is computed following the prescription described in
Section 5 of BC04, assuming observations with the two
Michelson-like observables.
For a given distance and observation time $T$, 
the SNR is determined by $M$, $m$ and
$e_f$. We sample $M$ between $10^{3.5}\msun$ and $10^{6}\msun$ at
intervals of ${\rm log}M=0.5$, $m$ on a grid of $0.1\msun$ between
$0.1\msun$ and $1.4\msun$, and we consider four values for the final
eccentricity: $e_f=0$, 0.1, 0.2, 0.3. The remaining parameter that
affects the SNR is the duration of the observation $T$; here we
consider $T = 3$ and 5~yr.

In figure~\ref{fig2} we show examples of the expected GW signal and
of the  eccentricity and the SNR evolution as a function 
of the observation time. We
show, in particular, the characteristic amplitude $h_{c,n}\propto
\dot{E_n}/\dot{f_n}$ -- here $\dot{E_n}$ is the energy emitted per
unit time in the $n$-th harmonic with observed frequency $f_n =
n\nu+\dot{\gamma}/\pi$, where $\gamma$ is a precession angle (see
BC04) and $\dot{f_n}$ is the time derivative of $f_n$ -- of the first
five harmonics as a function of frequency and the evolution of
eccentricity of the fiducial systems (for $e = 0$ only the harmonics 
corresponding
to $n=2$ contributes to the signal). Notice that EMRIs characterised by $e=0$
produce larger SNR than those with $e\ne0$. In fact the SNR
accumulated in the last few years of the life of a source decreases
with increasing eccentricity due to the steep increase of $\dot{f_n}$
as a function of $e$, and the increase of the {\it LISA} noise at
frequencies above $\approx 5$ mHz. Circular EMRIs can also be detected
at a fiducial distance of 100 Mpc with a SNR$\simgt 30$ several tens
of years before coalescence/disruption. This is not the case for
eccentric systems, because the peak of their emission is moved
progressively away from the ``sweet-spot'' of the {\it LISA}
sensitivity window. For systems characterized by a non-negligible
eccentricity at disruption $e\sim 0.1$, those with $m\sim0.5\msun$
produce the highest SNR.

At this stage we still do not have a sufficiently detailed
understanding of any end-to-end analysis algorithm for the detection of
EMRIs, and therefore it is impossible to adopt a precise value for the
SNR threshold at which detection can be achieved. Here we
will adopt SNR$= 30$, following Gair et al. (2004), a value consistent
with the performance of the very first implementation of algorithms
being explored within the Mock {\it LISA} Data Challenges (e.g. Babak
et al. 2008; Gair et al. 2008a,b; Cornish 2008).  After setting the 
detection threshold, it is
straightforward to compute the maximum luminosity distance $D_{\rm
MAX}$ at which a source can be detected as a function of masses and
eccentricity (for a given observation time).  
Contour plots of $D_{\rm MAX}$ in the $(M,m)$ plane and
for $e_f=0, 0.2$ are shown in figure \ref{fig5} assuming $T = 3$ and
5~yr. It is clear that circular binaries are
likely to be detected at much larger distances than systems 
characterised by $e_f\ne 0$. 
In the case $e_f=0$ there is a small region of parameter space 
for which the WD is
disrupted before the last stable orbit  
and is observable up to $D_{\rm MAX}\simeq 200-250$ Mpc. 
If we assume that a WD disrupted during the
final plunge would also yield a detectable electro-magnetic burst,
then {\it LISA} could probe even deeper, possibly reaching $D_{\rm
MAX}\simeq 450$ Mpc. As a general trend, for fixed
source masses the volume of the universe surveyed by {\it LISA}
decreases with increasing $e_f$. For example, if $e_f=0.2$, the
maximum distance at which {\it LISA} can detect a MBH-WD binary 
leading to WD disruption before the last stable orbit is $\approx$150 Mpc. 
$D_{\rm MAX}$ obviously
increases with observation time, however, even with only $T= 1$ yr
{\it LISA} can reach $D_{\rm MAX} \gsim 100$ Mpc. We will use the  values of
$D_{\rm MAX}$ derived in this section to obtain estimates of the LISA detection
rates of GWs from MBH-WD leading to the disruption of the WD in Section 
\ref{s:rate}. The detection rate scales as $D_{\rm MAX}^3$ and any 
uncertainty in the determination of the maximum distance will be 
reflected accordingly in the rates.

\subsection{The role of black hole spin}

\begin{figure}
\centerline{\psfig{file=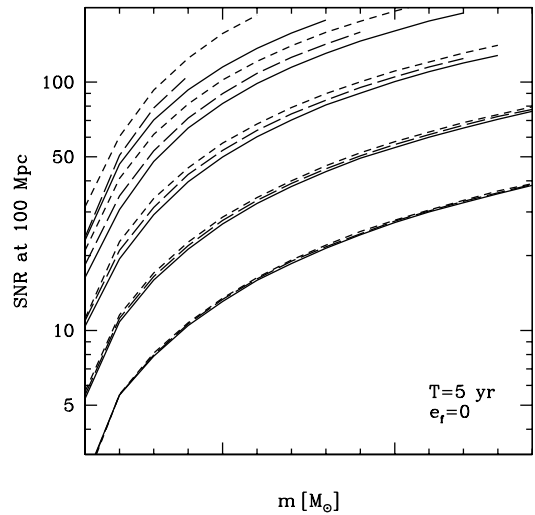,width=75.0mm}}
\caption{SNR obtained with 5 yr of coherent integration before
the maximum between $R_{\rm td}$ and $R_{\rm LSO}$. 
The WD is assumed to orbit a Kerr MBH with $a=0.99$ either in an 
equatorial prograde (solid lines) or retrograde (short--dashed
lines) orbit, or a Schwarzschild MBH (long--dashed lines).  
Each series of curves (solid, long--dashed or short--dashed) refers, 
from bottom to top, to WDs orbiting around a MBH
with $M=10^{3.5},10^4,10^{4.5}, 10^5,10^{5.5}\msun$.\label{fig6}}
\end{figure}
\begin{figure}
\centerline{\psfig{file=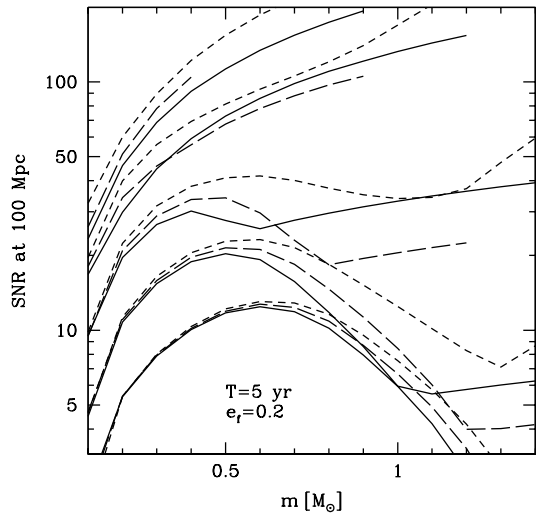,width=75.0mm}}
\caption{Same as figure \ref{fig6} but for systems with 
non negligible final eccentricity. Line--style and curve sequences as
in figure \ref{fig6}.\label{fig6b}}
\end{figure}
\begin{figure}
\centerline{\psfig{file=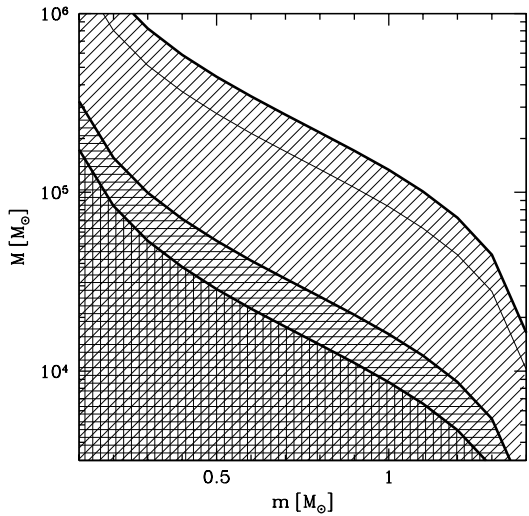,width=75.0mm}}
\caption{The mass parameter space of MBH-WD binary systems 
The thick solid lines mark the combination of MBH and WD mass 
for which the tidal disruption radius $R_{\rm td}$ 
equals the radius of the last stable orbit $R_\mathrm{LSO}$. From top to bottom
they refer to: a prograde orbit around a Kerr MBH with $a=0.99$, an
orbit around a Schwarzschild MBH, and a retrograde orbit around a Kerr
MBH with $a=0.99$. The shaded area(s) below the curves represent the portion
of the $M-m$ plane for which WD disruption occurs before the LSO and an
electro-magnetic counterpart is expected.
For comparison, the thin
line represents the mass combinations for which $R_{\rm td}$ equals
the radius of the horizon ($2M$) for a Schwarzschild
MBH. \label{fig7}.}
\end{figure}

The previous results were obtained assuming a Schwarzschild MBH. If
MBHs are rapidly rotating, as our present astrophysical understanding
suggests (e.g. Shapiro 2005), then relativistic spin-orbit coupling
affects the evolution of the binary and the resulting GW signal. To
quantify the impact of spins on the range of MBH-WD systems observable
with {\em LISA} with respect to the non-spinning case considered in
the previous Section we consider two extreme cases: in both instances 
the MBH is 
almost maximally rotating (the spin parameter is set to $a=0.99$), the 
WD orbit is equatorial, but in one case it is prograde and in the 
other it is retrograde.
The MBH spin affects the signal -- and ultimately the detection rate -- in 
two ways: (i) the strength of the emitted gravitational radiation, and (ii)
the distance at which a WD is tidally disrupted. The emitted GW
amplitude is larger for prograde orbits and so is the SNR (see
figures \ref{fig6} and \ref{fig6b}). This is a zero-to-$ 40\%$ 
effect increasing with the
BH mass, and favors the detection of prograde orbits.  However, the location
of the last stable orbit also shifts: for an equatorial prograde orbit, 
the LSO is at $R
= M$; in this case the WD is either disrupted during the inspiral
phase or simply plunges into the MBH horizon without producing any
electro-magnetic signature.  For an equatorial retrograde orbit, the
LSO is located at $R = 9M$; in this case the region where the WD could
be disrupted during the final plunge before entering the BH horizon is
much wider. We found that this LSO shift has a small influence on the
SNR (the essential point is that radiation generated when the orbital
frequency is close to the LSO, whether it is located at $M$ or $9M$,
is detected in a region of the {\it LISA} observational window where
the sensitivity is already compromised by the fact that the instrument
arm-length is significantly longer than the typical GW wavelength; see
figure \ref{fig2}), however, it strongly affects the portion of the
$M-m$ parameter space in which the WD tidal disruption occurs outside
the LSO, as shown in figure \ref{fig7}. The prograde LSO for a MBH
with $a=0.99$ is located at a much smaller radius, increasing the
maximum MBH mass that allows WD disruption {\it before} the final
plunge. For example, a 0.5$\msun$ WD is disrupted outside the LSO of a
Schwarzschild MBH of $M<5\times10^4\msun$; however, if the same WD is
in an equatorial prograde orbit around a Kerr BH with $a=0.99$, it is
disrupted outside the LSO if $M<5\times10^5\msun$. In this latter case
disruption caused by more (and more massive, i.e. detectable to larger
distances) MBHs is observable both in the gravitational wave window
and in the electro-magnetic band, yielding a much higher event rate.
For retrograde orbits, the LSO is larger -- $R_{\rm LSO}\approx 9M$ --
and the WD is disrupted by MBHs of smaller masses (see figure \ref{fig7})
. However, WDs will be captured in random direction with respect 
to the MBH spin  axis, and the detection rate will tend to increase on 
average, since what it is gained in 
terms of SNR and of available mass parameter space because of the prograde 
orbits is much more than what it is lost because of the retrograde ones. 
We therefore use as a reference for rate estimations the non spinning 
MBH case, as this sets a lower limit to the detection rate.

\section{Detection rate}
\label{s:rate}

We can now address the {\it LISA} detection rate of inspirals signals from 
MBH-WD binaries in which the WD is disrupted before falling into the 
MBH horizon. 
The detection rate $\Gamma$ can be computed via 
\be
\Gamma=\int dm \int dM\frac{d^2n}{dMdm}V(M,m)\,,
\label{rate}
\ee 
where $V(M,m)=4/3\pi D_{\rm MAX}^3(M,m)$ is the observable volume 
-- here we consider an Euclidean approximation, due to the low redshift of the
sources covered by the instrument, see Section 2 and figures~\ref{fig5}, ~\ref{fig6} and ~\ref{fig6b} -- 
and  
\be
\frac{d^2n}{dMdm}=\frac{dn}{dM}\times\frac{d\Gamma}{dm}\bigg|_M\,.
\label{fact}
\ee 
The detection rate depends sensitively on two rather unknown
quantities: the local number density of MBHs per unit mass $dn/dM$,
and the WD inspiral rate per unit WD mass $d\Gamma/dm|_M$, given the
MBH mass. The factorization in equation (\ref{fact}) is based on the simplifying
assumption that the inspiral rate is a function of the MBH mass only,
i.e., we are assuming that 
MBHs with the same mass are embedded in similar
stellar cusps.

At present there are no stringent constrains on $dn/dM$ in the mass
range of interest ($M \simlt 3\times10^5\msun$). The MBH mass function
was derived by Aller \& Richstone (2002) for $M>10^6\msun$ by
combining the galaxy luminosity function with the Faber-Jackson and
the $M-\sigma$ relations. The slope of the faint end of the luminosity
function is, however, controversial. Some works suggest that in
clusters and groups of galaxies it can be as steep as $L^{-1.8}$ (where $L$ is
the luminosity) for
magnitudes brighter than -18 (e.g. Wilson et al. 1997; Sabatini et al. 2003;
Gonzalez et al. 2006). In contrast, Trentham \& Tully (2002) find a
scaling of $L^{-1.2}$ in the magnitude range [-18,-10]. The
lack of information on the faint-end slope of the luminosity function
implies an uncertainty of a factor of $\sim 10$ on the number density
of MBHs in the range $10^4\,\msun \simlt M \simlt 10^5\,\msun$, even
assuming that the Faber-Jackson and the $M-\sigma$ relations can be
extrapolated to such small galaxies. This highlights the role that
{\it LISA} observations could play in providing information on the demographics
of MBHs in this mass range that has been proven difficult to probe so far,
and on their host galaxies.

Estimates of the WD inspiral rate onto a MBH embedded in a stellar
cusp are even more uncertain and they span several orders of
magnitude. Let us define the total rate for a single MBH as
\be
\Gamma|_M=\int \left.\frac{d\Gamma}{dm}\right|_M dm\,.
\ee. 
We assume for simplicity that the
differential rate is $d\Gamma/dm|_M=\Gamma|_M\, {p}_{\rm WD}$, where
$p_{\rm WD}$ is the WD mass density function   
taken from Madej, Nalezyty \& Althaus (2004).  In dense environments 
like compact star
clusters or dwarf galaxy nuclei (where we expect to find MBHs in the
mass range of interest), MBH-WD binaries can be generated primarily 
by (i) two body
scattering diffusion of stars into the MBH loss cone, and (ii) tidal
break-up of a stellar binary followed by the capture of one of the
components.  The former is likely to generate EMRIs with
non-negligible final eccentricity, while the latter is likely to
generate EMRIs with $e \approx 0$ (Miller et al. 2005; see
Amaro-Seoane et al. 2007 for a review).  Estimates of the capture rate
due to the two body scattering processes have been typically carried
out for a fiducial MBH mass $\sim 10^6\msun$, and the results then
rescaled to other MBH masses.  Hils \& Bender (1995) and Ivanov (2002)
find rates in the range $5\times 10^{-9}< \Gamma|_M <2\times10^{-8}$
yr$^{-1}$. Ivanov (2002) suggested a scaling $\propto M^{-1}$. Results
obtained by Sigurdsson \& Rees (1997) span several orders of
magnitude, depending on the radial density profile of the stellar
distribution. For typical observed power--law profiles 
$\rho_{\rm pl}\propto r_{\rm pl}^{-1.5, -2}$,
they suggest $10^{-10}< \Gamma|_M <5\times10^{-7}$ yr$^{-1}$.  Hopman
\& Alexander (2006a), taking into account mass segregation, give
$\Gamma|_M\approx 3\times 10^{-8}$ yr$^{-1}$ for a Milky Way MBH. This
number can be larger by a factor $\approx 10$ if resonant relaxation
is efficient (Hopman \& Alexander 2006b). They also estimate a scaling
$\propto M^{-0.25}$. Monte Carlo simulations performed by Freitag
(2001) give rates as high as $5\times10^{-6}$ yr$^{-1}$ for a Milky
Way MBH, and Gair et al. (2004) infer a mass scaling $\propto M^{3/8}$
on the basis of dynamical friction arguments.  On the other hand there
are currently no detailed calculations of the rate of EMRIs generated
through tidal break-up,
but there are clues suggesting that it could even
exceed the rate yielded by standard relaxation processes  
(Miller et al. 2005; Alexander 2007). 

It is clear that the theoretical understanding of the
magnitude and mass scaling of $\Gamma|_M$ is quite poor, which
highlights once more the potential impact of {\it LISA} observations
of this class of objects. The actual rate for $M < 10^6\msun$, which
is of interest here, depends both on how the rates scale with $M$ and
on how the properties of the nuclear cluster scale with $M$.
Observational data for the small dense clusters associated with low
mass nuclei are sparse.  If the observed inverse correlation between
$M$ and the stellar density in the nucleus holds for these lower
masses, then we expect the rates to be 3--10 times larger
($10^{-7}$-$10^{-6}$ y$^{-1}$) with respect to a $10^6\msun$ MBH,
taking the overlap of recent estimates and scaling, which ultimately
favors detection rates at the upper end of the range we will estimate
below.  It is also possible that the stellar zero-age mass function is
systematically skewed to higher masses in galactic nuclei, in which
case the median WD mass may be higher than estimated from the field
mass function, perhaps as high as 0.7--$0.8 \msun$, and the ratio of
WDs to current main sequence stars would be significantly higher.
This may enhance the rate by another factor of several due to stronger
scattering. On the other hand, mass segregation effects might lower
the rates, as the mean stellar mass in the nucleus would also be
significantly higher.

\begin{table}
\begin{center}
\begin{tabular}{c|cccc}
\hline
$$ & \multicolumn{4}{c}{Detection rates [yr$^{-1}$]}\\
\hline
$$ & \multicolumn{2}{c}{pessimistic} & \multicolumn{2}{c}{optimistic}\\ 
$$ & $e_f=0$ & $e_f=0.2$ & $e_f=0$ & $e_f=0.2$\\
\hline
   Schw no plunge&     0.03&    0.01&    21&   4\\
   Schw plunge&        0.15&    0.03&    71&   21\\
   Kerr prograde&      0.45&    0.40&    163&  124\\
\hline
\end{tabular}
\end{center}
\caption{Detection rates (events/yr) considering a 5 yr signal integration 
under different assumptions for MBH spins and WD orbits. 
The 'Schw no plunge' case assumes an event 
to be detectable both in the gravitational and electro-magnetic bands 
only if the WD tidal disruption occurs outside 
the last stable orbit, while in the 'Schw plunge' case we assume a 
visible electro-magnetic counterpart also if disruption 
takes place during the final plunge. The pessimistic and optimistic
scenarios are described in the text.}
\label{tab:1}
\end{table}

The detection rates of WD-MBH binaries leading to the tidal disruption
of the WD companion derived using equation (\ref{rate}) lie in the range
$0.01\,\mathrm{yr}^{-1} \simlt \Gamma \simlt 100$ yr$^{-1}$ and are
summarised in table \ref{tab:1}. Optimistic estimates are calculated
assuming the WD disruption rate quoted by Freitag (2001) with the Gair
et al. (2004) mass scaling coupled with a steep faint end of the MBH
mass function $dn/d{\rm log}M\propto M^{-1}$ (e.g. Gonzalez et
al. 2006); these assumptions lead to $\sim 5-150$ events per year,
depending on MBH spin, WD eccentricity, and the assumption that
an electro-magnetic counterpart exists when the WD is disrupted outside the
black hole horizon, and not simply before the last stable circular orbit.
Pessimistic estimates rely on WD disruption rates derived by
Hopman \& Alexander (2006b), in which no mass segregation or resonant
relaxation effects are taken into account, coupled with a flat 
extrapolation of the MBH mass function ($dn/d{\rm log}M={\rm cost.}$, 
Aller \& Richstone 2002), and are in the range $\sim 0.01-0.5$ yr$^{-1}$. It is
then likely that {\it LISA} (assuming a five years operation lifetime)
would observe several of these MBH-WD sources. Detection
rates for some fiducial cases are summarised in Table\ref{tab:1}.
Note that the presence of binaries with $e \approx 0$ could be decisive 
for successful {\it LISA} detections, because circular systems can 
be detected to significantly larger distances (see figure \ref{fig5}) 
and the associated rates are much higher, as shown in table \ref{tab:1}.

\section{Electromagnetic counterparts}
\label{e:EM}

In this section we model the electro-magnetic signature of a WD
disrupted {\it outside} the last stable orbit and discuss the 
prospects for observing such events. Whether a disruption during the final
plunge would lead to any distinct electro-magnetic signal, is still an
open question that would require full relativistic hydrodynamical
simulations, that go beyond the scope of this paper and will not be 
addressed here. 

The prompt electro-magnetic signature of the disruption of a 
WD from a MBH consists of an
energetic flare that appears when accretion begins and lasts while
accretion fuel is available. Since the orbit decays very quickly via
gravitational radiation, the only limitation to the accretion rate is
the Eddington limit. Thus, the bolometric luminosity can be written as
$L_{\rm bol}=1.3\times 10^{43}\, (M/10^5\,\msun)~{\rm erg~s}^{-1}$.
The corresponding accretion rate is 
$\dot{M} \approx 2\times 10^{-3}\, (\epsilon/0.1)^{-1}\,
(M/10^5\,\msun)~{\rm \msun~yr}^{-1}$, where $\epsilon$ 
is the efficiency of conversion of the rest energy of the accreting matter to
radiation. At this rate, a CO WD with a mass of $0.55~\msun$ will be
consumed in about 300 years. Flares with such luminosities can be
easily detected with present-day X-ray observatories out to the
distance of the Coma cluster ($\sim 100$~Mpc). By analogy with the
results of Rantsiou et al. (2008) who compute the tidal disruption of
a neutron star by a $10~\msun$ black hole, we expect that a portion of
the post-disruption debris will become unbound and expand radially
outwards with a morphology resembling an arc. Photoionization of this
arc of debris by the accretion-powered X-rays and UV light will lead
to the emission of optical lines. The late-time afterglow spectrum of
the debris makes up a very distinct signature of such an event.

In order to calculate the emission-line strengths from the
photoionized debris, we first estimate how its physical properties
will evolve with time. We assume that the debris arc contains a small
fraction, of order 10\%, of the mass of the WD and it is
launched from the tidal disruption radius at the escape speed. As the
debris expands, it remains primarily in the original orbital plane but
also expands perpendicularly to it at the speed of sound $c_s$.
Because of photoionization, the excitation temperature of the
electrons in the debris remains relatively steady at $T\sim{\rm
a~few}\times 10^4$~K, thus $c_s\approx 10\,(T/10^4\, {\rm
K})^{1/2}\;{\rm km~s}^{-1}$ (this is based on the photoionization
calculations that we describe in the next paragraph). 
As fiducial case we consider 
a $0.55~\msun$ CO WD disrupted by a
$10^5~\msun$ MBH. This WD mass represents the peak of
the mass distribution found by Madej, Nale\'zyty \& Althaus 2004), 
and  such WD-MBH combination would lead to the disruption of the star 
for mildly spinning MBHs and prograde orbits, and the GW precursor would 
be detectable well beyond 200 Mpc. 
We also consider briefly the emission-line spectrum from the
disruption of a $0.4~\msun$ He WD, which is a less likely
event. The tidal disruption radius of the CO WD is
$R_{\rm td}=5.3\times 10^{10}$~cm and the dynamical time at this radius
is $t_{\rm dyn} (R_{\rm td}) = 2.4$~s.  Under the above assumptions, at
$t\gg t_{\rm dyn} (R_{\rm td})$, the radius of the debris increases with
time as $r_d(t)\propto t^{2/3}$ and reaches $9\times 10^{13}$~cm after
approximately 31 hours. Accordingly, the density of the debris
evolves as $\rho_d(t)\propto t^{-7/3}$ and the column density through
the debris (along the orbital plane) decay as $t^{-5/3}$. 
The solid angle subtended by the arc of debris to the continuum source 
{\it increases} with time as $t^{1/3}$ because the expansion rate at the
speed of sound perpendicular to the plane of the orbit is slightly
faster than the expansion in the orbital plane.

To compute the ionization structure and emission properties of the
debris, we used the photoionization code Cloudy (Ferland et al. 1998),
which computes the thermal and ionization balance of the debris in
detail, taking into account a wide variety of physical processes and
emission mechanisms. The ionization level of the gas is described by
the ionization parameter $U_H=Q_H/(n_H r^2_d c)$, where $Q_H=\int_{\rm
1\;Ry}^{\infty}(L_{\nu}/h\nu)\; d\nu$ is the emission rate of hydrogen-ionizing 
photons by the central accreting MBH, $n_H$ is the
hydrogen number density in the debris and $r_d$ is its distance from the
central source of ionizing photons.  In this particular case, we must
re-define the ionization parameter because hydrogen is very sparse in
the photoionized debris. We choose oxygen as the reference element
instead because it makes up most of the mass in the debris and because
its first ionization potential happens to be nearly identical to that
of hydrogen. The two ionization parameters are related by $U_O =
(m_H\, A_O/m_O) U_H$, where $m_H$ and $m_O$ are hydrogen and oxygen
masses, respectively, and $A_O=16$ is the mass number of oxygen.
Thus, $U_O=U(t)\propto L(t)/[\rho_d(t) r^2_d(t)]$.  We make the assumption
that the accretion rate, hence the ionizing luminosity, remains steady
at the Eddington limit in the immediate aftermath of the disruption,
implying that $U(t)\propto t$. The composition of the debris is a
crucial ingredient of the calculation. Following Madej et al. (2004),
we assume that the mass of the CO WD consists of 67~\% of O, 32~\% of
C, 1~\% of He, 0.001\% of H, and other metals in the mass fractions
observed in the Sun (Anders \& Grevesse 1989). We note that this
composition is consistent with WD astroseismology results (Metcalfe
2003; Staniero et al. 2003). We adopted the Mathews \& Ferland (1987)
model for the spectral energy distribution of the ionizing continuum,
which is motivated by observations of quasars.  This model comprises a
hard power-law in the X-ray band and an excess of UV photons (a "UV
bump").

\begin{figure}
\centerline{\psfig{file=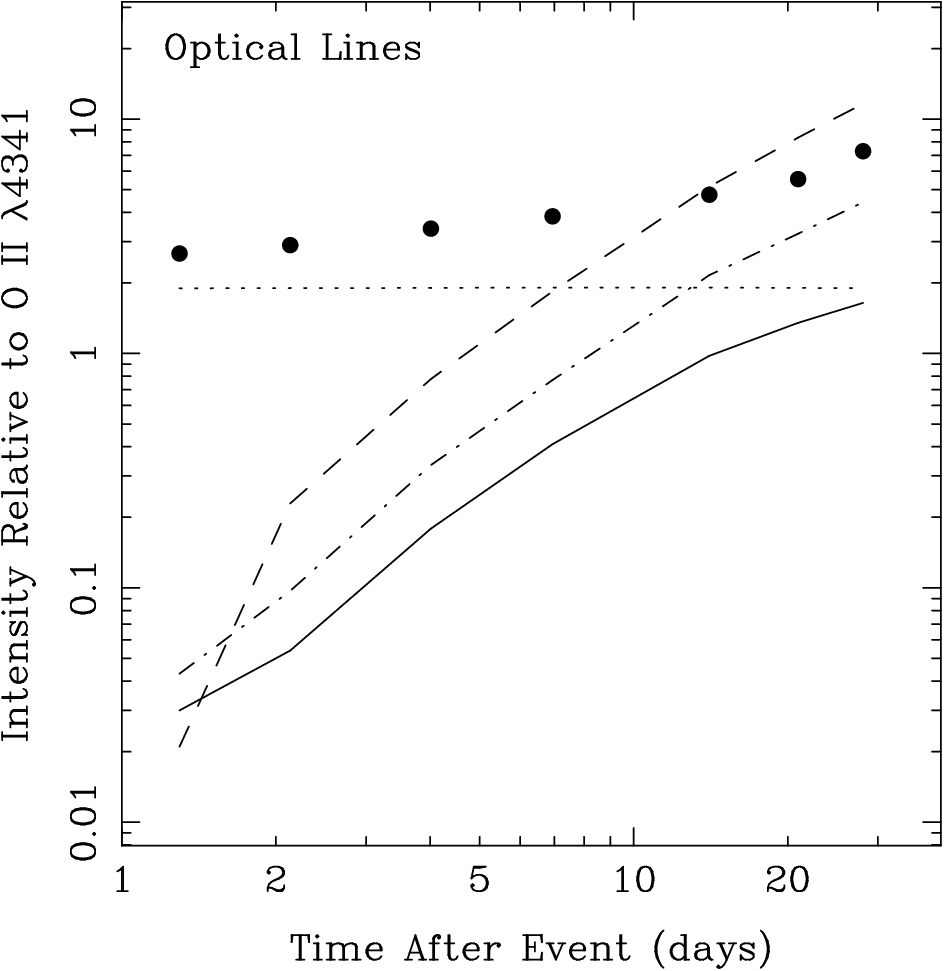,width=75.0mm}}
\caption{Time evolution of the intensities of selected, strong, optical
emission lines relative to \lion{O}{2}{4341}. Solid line --
\lion{O}{1}{4368}; dashed line -- \lion{C}{1}{8727}; dotted line --
\lion{O}{2}{4651}; dot-dashed line -- \lion{Ca}{2}{3934}. The large
filled circles show the evolution of the luminosity of the reference
line, \lion{O}{2}{4341}, in units of $10^{36}~{\rm
erg~s}^{-1}$.\label{fig_opt}}
\bigskip
\centerline{\psfig{file=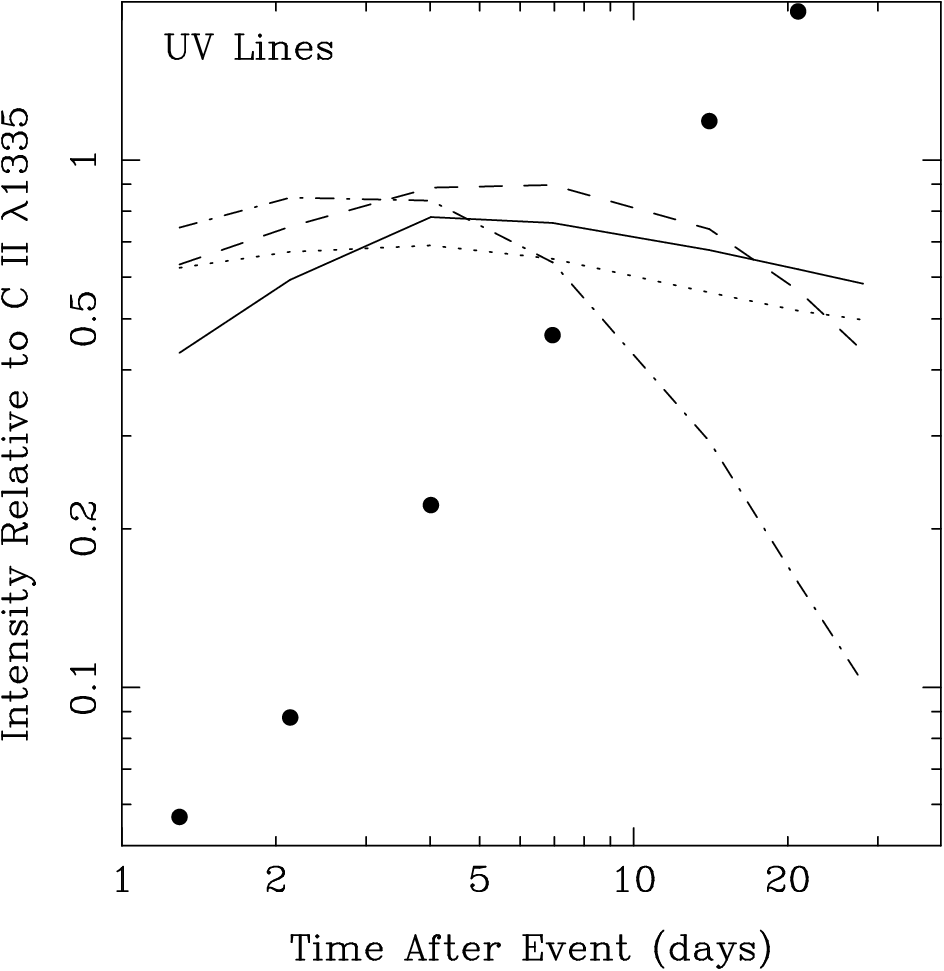,width=75.0mm}}
\caption{Time evolution of the intensities of selected, strong, ultraviolet
emission lines relative to \lion{C}{2}{1335}. Solid line --
\lion{C}{4}{1551}; dashed line -- \lion{C}{1}{1656}; dotted line --
\lion{Si}{3}{1207}; dot-dashed line -- \lion{C}{1}{1248}. The large
filled circles show the evolution of the luminosity of the reference
line, \lion{C}{2}{1335}, in units of $10^{36}~{\rm erg~s}^{-1}$.\label{fig_uv}}
\end{figure}

The photoionization calculations predict an emission spectrum of C and
O recombination lines, as one would expect. We have calculated models
at 1, 2, 4, 7, 14, 21, and 28 days after the event, which show that
the ionization parameter increases from $\log U_O=-5.8$ at $\sim1$~day
after the event to $\log U_O=-4.4$ at 4 weeks after the event, as the
debris expands and thins out.  The strongest oxygen and carbon lines
in the optical spectrum are \lion{O}{2}{4341}, \lion{O}{1}{4368},
\lion{O}{1}{8446}, \lion{O}{2}{4651}, and \lion{C}{1}{8727}. In
addition the \ion{Ca}{2}~H\&K doublet ($\lambda\lambda$3934,3969) and
the \lion{N}{1}{3466} line appear to be strong. The evolution of the
optical lines with time is shown in Figure~\ref{fig_opt}. In the
near-UV part of the spectrum, the strongest lines are from carbon,
e.g., \lion{C}{2}{1335} \lion{C}{1}{1656}, \lion{C}{1}{1248},
\llion{C}{4}{1548,1551}, and silicon, namely \lion{Si}{3}{1207},
\lion{Si}{2}{1263}, \llion{Si}{4}{1394,1403}. The evolution of the
near-UV lines with time is shown in Figure~\ref{fig_uv}. The behavior
seen in different lines, i.e., the fact that specific lines rise with
time while other fall, constitutes a diagnostic test for identifying
this type of event.

The behavior of the spectrum of helium-rich debris from the disruption
of a $0.4~\msun$ WD is qualitatively similar. The strong optical
emission lines are primarily from helium, and include
\lion{He}{1}{5876}, $\lambda$4026, and $\lambda$4471 and
\lion{He}{2}{3203}, $\lambda$4686, and $\lambda$5412. The \ion{He}{2}
lines are predicted to rise with time relative to the \ion{He}{1}
lines. In the near-UV band, the strong lines are the same as those of
the CO WD case, with the addition of one strong helium line,
\lion{He}{2}{1640}. The characteristic behavior we predict in the UV
band is that the \lion{C}{2}{1335}, \llion{Si}{4}{1394,1403} and
\lion{Si}{3}{1207} lines decline with time relative to
\lion{He}{2}{1640}.

Our emission-line luminosity estimates are subject to uncertainties in
the distribution of the post-disruption debris and the evolution of
ionizing luminosity with time (we have assumed the ionizing luminosity
to be steady during the first month after the event). Nevertheless, an
interesting feature of our predicted luminosity evolution is the
increase in the luminosity of some of the strong emission lines with
time. This is a consequence of the fact that the excitation rate of
these lines increases as the density of the debris drops allowing the
ionizing photons to penetrate deeper into the debris.  Taking these
luminosity estimates at face value, we expect that the optical
emission lines should be detectable out to the distance of the Virgo
cluster ($\sim 16$~Mpc) with a few-hour exposure on a 4m-class
telescope, and well beyond the Coma cluster with a few hour
exposure on a 30m-class telescope. The UV lines should be detectable
out to the distance of the Virgo cluster with a telescope of the class of 
the {\it Hubble Space Telescope}. The capabilities of present and near
future observatories should therefore allow the identification of the
electro-magnetic counterpart associated to the LISA detections gravitational 
waves of MBH-WD within the relevant parameter range.

\section{Summary}
\label{e:summary}

In this paper we have considered a class of {\it LISA} sources that have
not been explored so far, consisting of a WD inspiralling onto a MBH
in the mass range $\sim 10^4-10^5\,\msun$. These sources are of
particular interest because the gravitational wave signal produced during
the inspiral can be detected with {\em LISA} and serves as a precursor for
an electro-magnetic flash -- likely detectable in X-ray, and at optical and 
ultra-violet wavelengths -- generated during the tidal disruption of the star 
and the subsequent distinctive accretion episode, which takes place for a 
considerable region of the mass parameter space.  Observations of the same 
source in the gravitational and electromagnetic band enable an independent 
and calibration-free determination of the Hubble parameter at low 
redshift $z\simlt 0.1$ (through the direct measurement
of the redshift {\em and} luminosity distance of the same source), studies 
of the faint end of the MBH mass function in (dwarf) galaxies, measurements 
of the mass distribution function of 
extra-galactic WDs and in particular of those in galaxy cores,
and provide a probe of the structure and equation
of state of WDs, and of dynamical processes in the cores of galaxies.
 
We have determined the range of WD and MBH masses that lead to a tidal
disruption before the final plunge, hence producing an
electro-magnetic counterpart to the GW emission.  In order to do so we have 
adopted a polytropic approximation of the WD mass-radius relation. In the future
it is important to explore more sophisticated models to establish the dependence
of the results presented here on this assumption. We have modelled the
expected GW signal using "analytical kludge" waveforms {\em a-la} 
Barack \& Cutler and have explored the dependence of the SNR on selected values
of the eccentricity and the MBH spin.
We found that the maximum distance at which {\it LISA} can observe
these events depends both on the MBH spin and the orbital
eccentricity, with circular prograde orbits around a highly spinning
MBH producing signals observable to larger distances. A typical
0.5$\msun$ WD orbiting a $5\times10^4\msun$ Schwarzschild MBH would be
observable up to $200$ Mpc, assuming a detection threshold 
equivalent to an optimal SNR of 30 in the two combined noise
orthogonal unequal-arm Michelson observables for 5 years of integration.

 Based on current (rather uncertain) estimates of the MBH
population and of the rates at which they capture WDs we have computed
the LISA detection rate and discussed its uncertainties.
Circular binaries yield a higher detection rate 
(by up to a factor of $\sim 3$) with
respect to eccentric ones, due to their higher SNR. A substantial
population of highly spinning MBH would enhance the detection rate,
since the last stable circular orbit of a WD inspiralling on a 
prograde orbit would be much
smaller. More massive (i.e. observable to further distances) MBHs
would enhance the detection rate. The rates that we obtain span almost
four orders of magnitude, from $0.01$ to $100$ events per year and are
listed in table \ref{tab:1}. Scaling arguments for the stellar
density in galactic centers suggest high WD disruption rates, up to
10$^{-6}$ yr$^{-1}$, in dwarf nuclei favoring the upper end of the detection rate
range quoted above. Assuming a minimum lifetime of 5 yrs, {\it LISA}
will likely see several of these events.

We have also modeled the electro-magnetic signature associated to the WD
disruption. Assuming disruption outside the last stable orbit, part of the WD
material will settle in an accretion disk powering a luminous X-ray
source that will be detectable up to $\sim 200$ Mpc. A small fraction
of the debris will become unbound forming a thin arc or annulus that
expands at nearly the escape velocity.  Photoionization of this
annulus by X-ray and UV photons produced by the accretion process will
result in several different (depending on the WD composition) emission
lines, with relative intensities changing on a timescale of a week as
the debris expands. Such lines will be observable well beyond 100 Mpc
with future 30m-class telescopes and provide a unique signature
of this kind of events. If disruptions during the final plunge also
result in observable electro-magnetic flares, the detection rate would
increase by a factor of $\sim 3$ (see table \ref{tab:1}) considering
Schwarzschild MBHs. However whether such a disruption would produce an
observable signal is an open question that we have not tried to address
here and deserves further investigation.

Based on the above results it is conceivable that several 
MBH-WD binary systems will be observed both in gravitational wave 
and electro-magnetic band at a distance up to a few hundreds of Mpc; 
however no events as well as hundreds of events are consistent
with our current understanding of the key physical processes.

The need to observe MBH-WD binaries in the electro-magnetic and gravitational 
window to maximise the science return raises the issue of {\em prompt alerts} 
generated by {\it LISA} to electro-magnetic observatories. So far alerts have 
been discussed only in the context of observations of massive-black hole 
binary systems ({\em e.g.} Kocsis et al 2006, 2007a,b; Lang \& Hughes 2008), 
but the scenario that we have discussed in this paper calls for a systematic 
study regarding extreme-mass ratio-inspirals in the relevant mass range. 
Unfortunately, the complexity of searches for EMRIs and the still limited 
understanding and maturity of end-to-end algorithms (e.g. Babak et al. 2008; 
Gair et al. 2008a,b; Cornish 2008) prevent at present a realistic study of 
such an important problem.

\section*{Acknowledgments}
Support for this work was provided by NASA through grant GO-10131 from the
Space Telescope Science Institute and through grant ATP NNG04GU99G.
Partial support was provided by a grant from the NSF, PHY 02-03046. A.S 
is supported by the CGWP at Penn State University. 
A.V is supported by the UK Science and Technology 
Facilities Council. A.V., M.E. and S.S. acknowledge support from 
the Theoretical Astrophysics Visitors' Fund at Northwestern 
University and thanks the members of the group for their warm 
hospitality during their stay. A.S. and S.S. thank the Aspen Center 
for Physics for hospitality.

{}

\end{document}